\documentclass[twocolumn]{webofc}

\usepackage[varg]{txfonts}   
\usepackage{hyperref}
\usepackage{url}
\hypersetup{colorlinks=true,citecolor=blue,urlcolor=blue,linkcolor=blue}
\begin{document}
\title{Multiple shape coexistence in the $N=Z$ $^{84}$Mo nucleus}
%
%

\author{\firstname{Tom\'as R.} \lastname{Rodr\'iguez}\inst{1}\fnsep\thanks{\email{tr.rodriguez@ucm.es}}
}

\institute{Grupo de F\'isica Nuclear, Departamento de Estructura de la Materia, F\'isica T\'ermica y Electr\'onica and IPARCOS, Universidad Complutense de Madrid, Plaza de Ciencias 1, E-28040, Madrid (Spain) 
          }

\abstract{The structure of the nucleus $^{84}$Mo has been studied using the projected generator coordinate method (PGCM) with the Gogny D1S interaction. The calculations incorporate a mixing of particle-number and angular-momentum projected intrinsic wave functions, defined over triaxial quadrupole degrees of freedom. This approach yields an excellent agreement with the scarce experimental data for this nucleus and several bands based on different shapes are predicted.
}
\maketitle
\section{Introduction}
\label{intro}
The appearance of two or more bands in the low-lying spectrum of certain atomic nuclei with distinctive collective behaviors is interpreted as evidence for nuclear shape coexistence. In the region of neutron-deficient nuclei around 
$N=Z$, shape coexistence has been identified through the observation of low-lying $0^{+}$ states, bands with varying moments of inertia, spectroscopic quadrupole moments, and/or decay properties (see Ref.\cite{PPNP_124_103931_2022} and references therein). Specifically, experimental data for the $^{80}$Zr nucleus ($N=Z=40$) and its surroundings indicate bands consistent with deformed ground states~\cite{PRL_59_1270_1987, PRL_124_152501_2020, PRC_85_041303_2012, PRL_95_022502_2005, PRC_80_031304_2009, PRC_56_2497_1997, PRC_65_051303_2002}. However, this region is challenging to study experimentally due to its proximity to the proton drip-line and is also difficult to investigate theoretically, as it involves at least the $pf$ and $gds$ shells for both protons and neutrons. Nevertheless, studies using mean-field calculations, shell model approaches, and more recently, ab initio calculations within the coupled-cluster formalism have been performed (see Ref.\cite{PRC_110_L011302_2024} and references therein) showing a shape-coexistent behavior. 

One of the most suitable methods for studying potential shape coexistence is the projected generator coordinate method (PGCM), as the wave functions of the nuclear states are combinations of intrinsic wave functions with different deformations. This formalism has been applied in this region with the Gogny interaction to investigate shape evolution in isotopic chains such as krypton\cite{PRC_90_034306_2014}, as well as to predict multiple shape coexistence in the $^{80}$Zr nucleus~\cite{PLB_705_255_2011}. In this work, we present novel PGCM calculations for the $^{84}$Mo nucleus $(N=Z=42)$. The document is organized as follows. Section~\ref{theory} provides a brief review of the underlying theory. The results, including energy surfaces and collective wave functions in the triaxial plane, along with excitation energies, are discussed in Sec.\ref{results}. Finally, a summary and outlook on future work are presented in Sec.\ref{summary}.
\section{Theoretical framework}
\label{theory}
\begin{figure*}[t]
\centering
\includegraphics[width=1.0\textwidth]{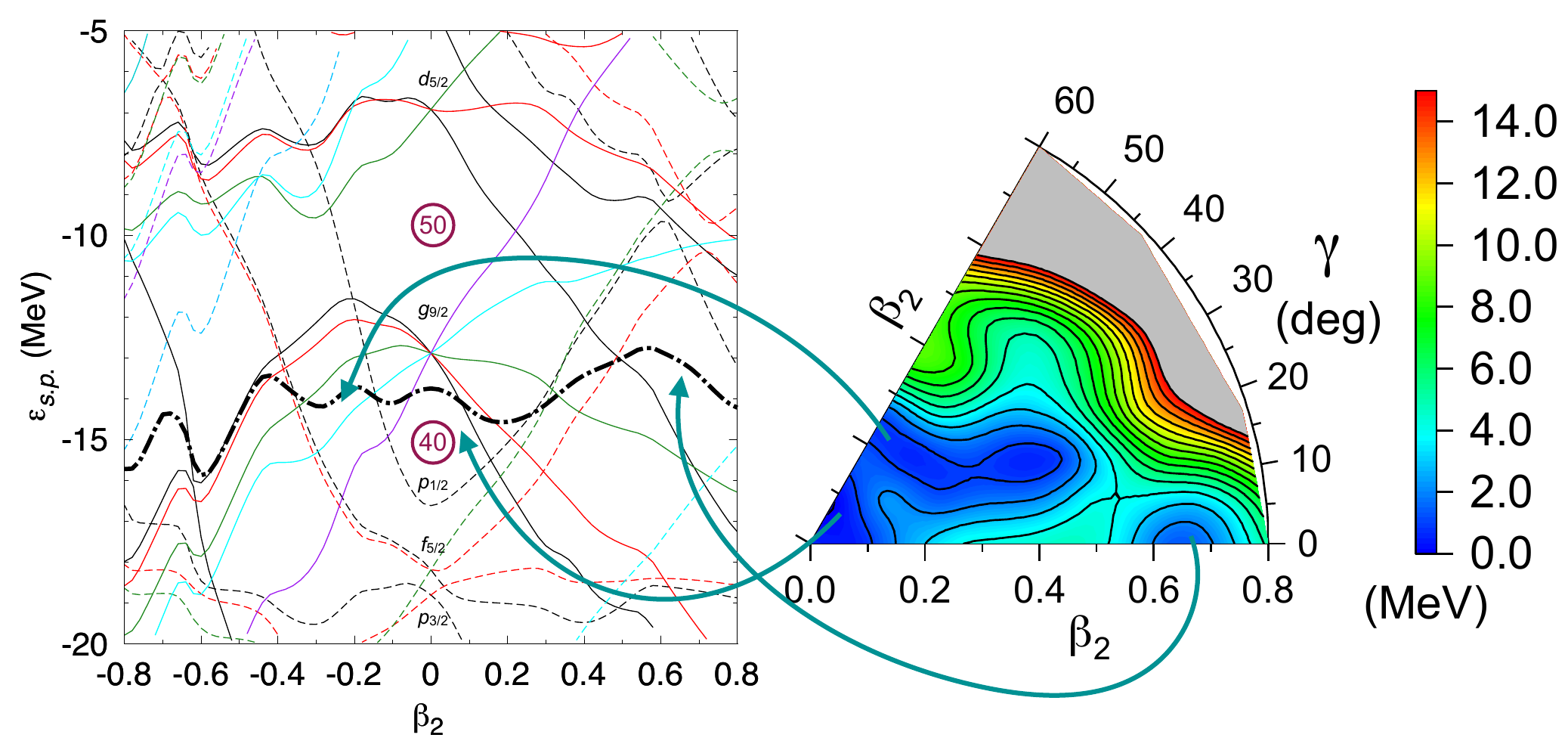}
\caption{(Right) Particle-number-projected energy surface in the triaxial plane; and, (left) single-particle-energies along the axial deformation (Nilsson-like levels, dashed, continuous, and dash-dotted lines represent negative and positive orbits, and the Fermi energy, respectively), calculated with the Gogny D1S energy density functional for $^{84}$Mo.}
\label{PNVAP_TES_Nilsson}      
\end{figure*}
\begin{figure}[b]
\centering
\includegraphics[width=1.0\columnwidth]{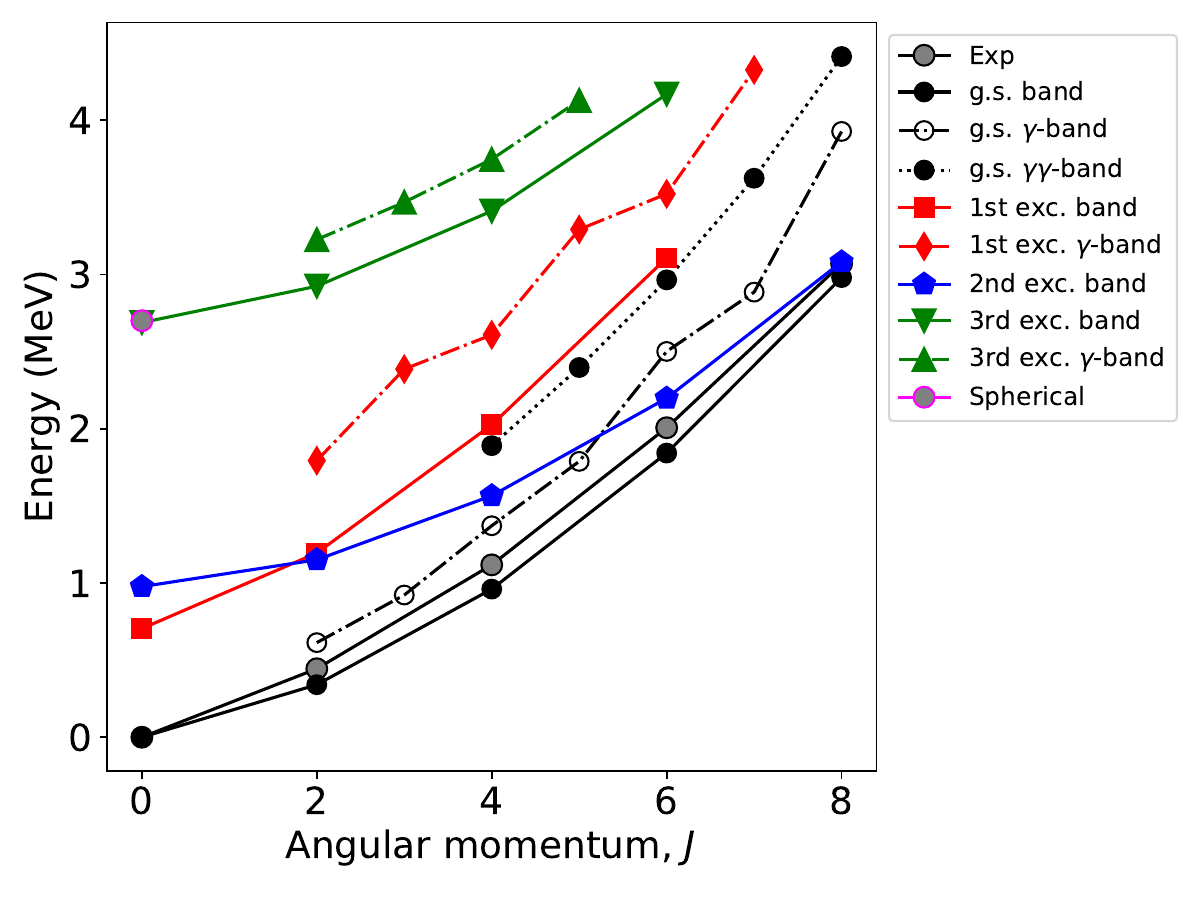}
\caption{Energy spectrum predicted by PGCM calculations with Gogny D1S energy density functional. Different bands are represented with the same symbols and colors to guide the eye. The experimental values~\cite{PRC_56_2497_1997,PRC_65_051303_2002} are also shown.}
\label{spectrum}      
\end{figure}
\begin{figure*}[t]
\centering
\includegraphics[width=1.0\textwidth]{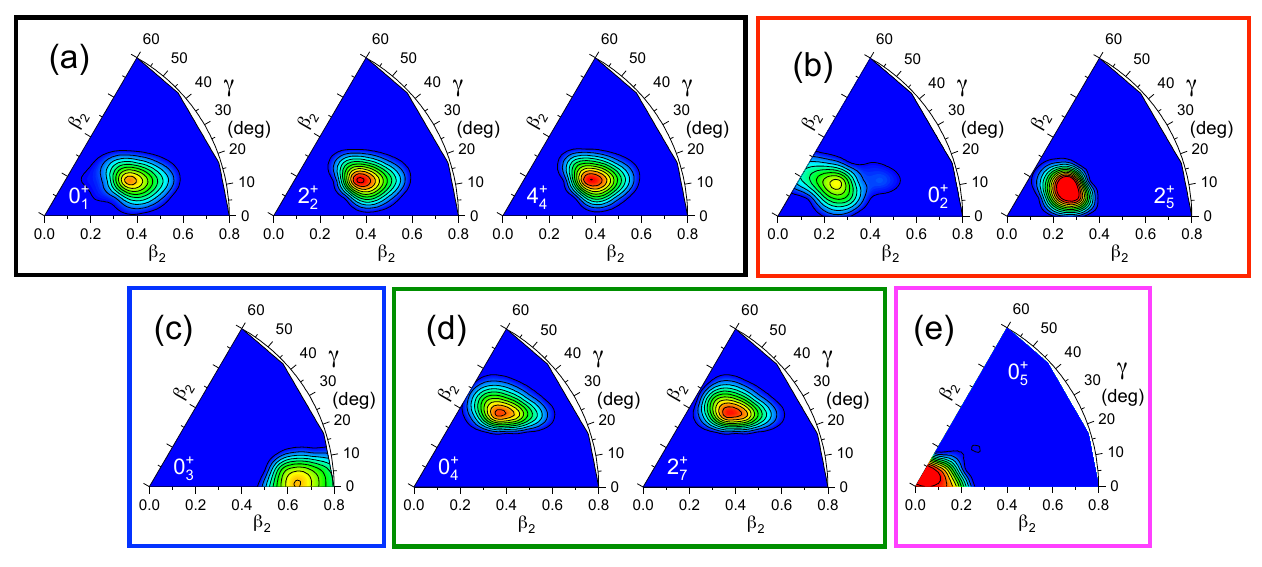}
\caption{Collective wave functions for the band-heads of the different bands obtained with the Gogny D1S-PGCM calculation for $^{84}$Mo. The color of the boxes coincides with the color code used to identify the bands in Fig.~\ref{spectrum}.}
\label{cwfs}      
\end{figure*}
The PGCM method, also know as symmetry conserving configuration mixing (SCCM) method, is based on the variational principle to approximate the nuclear states corresponding to a system of $Z$ protons and $N$ neutrons that interact, in the present implementation, with the Gogny D1S interaction. Such nuclear wave functions are defined as the linear combination of particle number and angular momentum projected Hartree-Fock-Bogoliubov (HFB) quasiparticle vacua, $|\phi_{(\beta_{2},\gamma)}\rangle$:
\begin{equation}
|JM;\sigma\rangle=\sum_{K,(\beta_{2},\gamma)}f^{J\sigma}_{K,(\beta_{2},\gamma)}|JMK;(\beta_{2},\gamma)\rangle.
\label{GCM_state}
\end{equation}
Here $J$, $K$ and $M$ are the angular momentum and the $z-$component in the body-fixed and laboratory frames, respectively, $\sigma=1,2,...$ sorts out the states for a given angular momentum, and $|JMK;(\beta_{2},\gamma)\rangle=\hat{P}^{N}\hat{P}^{Z}\hat{P}^{J}_{MK}|\phi_{(\beta_{2},\gamma)}\rangle$ are the particle-number and angular momentum projected wave functions (being $\hat{P}$ the projection operators~\cite{RingSchuck}). The intrinsic HFB vacua are obtained by minimizing the particle-number-projected HFB energy~\cite{NPA_696_467_2001} with constraints in the values of the triaxial quadrupole deformations parametrized by $(\beta_{2},\gamma)$~\cite{PRC_81_064323_2010}. Finally, the coefficients $f^{J\sigma}_{K,(\beta_{2},\gamma)}$ are obtained by solving the Hill-Wheeler-Griffin (HWG) equation that is deduced from minimizing the PGCM energy assuming Eq.~\ref{GCM_state} as the variational wave functions~\cite{RingSchuck,PRC_81_064323_2010}:
\begin{eqnarray}
\sum_{\xi'}\left(\mathcal{H}^{J}_{\xi\xi'}-E^{J\sigma}\mathcal{N}^{J}_{\xi\xi'}\right)f^{J\sigma}_{\xi'}=0&\nonumber\\
\mathcal{H}^{J}_{\xi\xi'}=\langle JMK;(\beta_{2},\gamma)|\hat{H}|JMK';(\beta'_{2},\gamma')\rangle&\nonumber\\
\mathcal{N}^{J}_{\xi\xi'}=\langle JMK;(\beta_{2},\gamma)|JMK';(\beta'_{2},\gamma')\rangle&
\label{HWG_eq}
\end{eqnarray}
In the above expression, $\xi\equiv\lbrace K,(\beta_{2},\gamma)\rbrace$ encodes $K$ and deformation parameters, and $E^{J\sigma}$ are the final PGCM energies that can be directly compared with the experimental data and/or other theoretical approaches.
\section{Results}
\label{results}
The formalism described above is applied now to the calculation of the $^{84}$Mo nucleus. A first insight of the collective structure of the nucleus is provided by the particle-number-projected energy as a function of deformation $(\beta_{2},\gamma)$. As shown in Fig.\ref{PNVAP_TES_Nilsson}, this energy reveals a structure with several minima: a spherical minimum, a triaxial minimum at $(\beta_{2},\gamma)\approx(0.4,20^{\circ})$ which is connected to rather degenerate deformations towards the oblate part and lower deformation $(0.25,30^{\circ}-60^{\circ})$, a highly deformed prolate minimum at $(0.65,0^{\circ})$, and a final depression at $(0.5,40^{\circ})$. This structure resembles the multiple shape coexistence observed in the $^{80}$Zr nucleus using the same method~\cite{PLB_705_255_2011}. As shown in that reference, the underlying Nilsson-type shell structure for this nucleus reveals that these minima correspond to Fermi level crossings at gaps in the single-particle energies. Specifically, the emptying (filling) of orbitals originating from the spherical $1p0f$ ($0g_{9/2}$ and $1d_{5/2}$) shells for both protons and neutrons generates this rich structure of minima in the energy surface.

This result suggests that the $^{84}$Mo nucleus is a strong candidate for exhibiting multiple shape coexistence, as the energy minima are close in energy. Furthermore, states with significant shape mixing could appear, given that these minima are separated by only modest energy barriers. Thus, the restoration of rotational symmetry and the configuration mixing within the PGCM framework is fully justified. The resulting energies from solving the HWG equation (Eq.~\ref{HWG_eq}) for each angular momentum value are shown in Fig.~\ref{spectrum}. Several rotational bands emerge, along with $\gamma$-bands (predominantly $K=2$) and $\gamma\gamma$-bands (predominantly $K=4$) associated with them. To interpret these bands in terms of intrinsic deformations, the collective wave functions (c.w.f.'s) of the band-heads are shown in Fig.~\ref{cwfs}. The remaining individual states within these bands exhibit c.w.f.'s similar to those of the band-heads (not shown), demonstrating the collective character of these states. In all cases, the maxima of the c.w.f.'s are concentrated at deformations where minima appear on the energy surface. Accordingly, the ground state band (band-head $0^{+}_{1}$, $\Delta J=2$) and its associated $\gamma$-band (band-head $2^{+}_{2}$, $\Delta J=1$) and $\gamma\gamma$-band (band-head $4^{+}_{4}$, $\Delta J=1$) are formed by triaxial deformations around $(\beta_{2},\gamma)\approx(0.4,20^{\circ})$; the bands built upon the $0^{+}_{2}$ ($\Delta J=2$) and $2^{+}_{5}$ ($\Delta J=1$) states show triaxial deformation around $(0.3,30^{\circ})$; a superdeformed prolate $\Delta J=2$ band is observed on the $0^{+}_{3}$ state with deformation $(0.65,0^{\circ})$; another $\Delta J=2$ band and its associated $\Delta J=1$ $\gamma$-band exhibit triaxial deformations around $(0.5,40^{\circ})$; and, finally, a spherical state, $0^{+}_{5}$, appears nearly at the same energy as the $0^{+}_{4}$ state. Therefore, after performing the full configuration mixing, a structure similar to that observed in the $^{80}$Zr nucleus is evident, where up to five $0^{+}$ states appear in the lower part of the spectrum, each associated with different shapes that correspond to the minima on the energy surface. In this case, the ground state is triaxial rather than prolate, as seen in the $^{80}$Zr isotope.

Finally, theoretical results are compared with the limited experimental data available for this nucleus. In Fig.~\ref{spectrum}, it can be seen that the excitation energies of the experimental ground-state band are well described by the theoretical results. However, this agreement could be considered “too good” given the limitations inherent in the PGCM implementation used here. In this study, excited states are not explored variationally as well as the ground state, so a stretching of the spectrum is expected. This stretching is reduced when time-reversal symmetry breaking terms, such as those obtained with rotating (cranking) intrinsic wave functions, are included~\cite{PLB_746_341_2015}. However, the theoretical energies without these terms are already below the experimental values up to an angular momentum of $J=8$, so including them would worsen the quantitative agreement with experiment. This may indicate two effects. Firstly, the experimental deformation might be overestimated; thus, the theoretically stretched energy of the $2^{+}_{1}$ state is small -and close to the experimental value- because of this excess of deformation. This occurs because angular momentum projection tends to shift the mean-field energy surface minima toward slightly larger deformations, and since the interaction is adjusted to reproduce certain experimental data at the mean-field level, the restoration of rotational symmetry produces these slightly larger deformations~\cite{PRL_94_102503_2005,PRC_91_044315_2015}. The second factor absent in these calculations is proton-neutron pairing~\cite{PPNP_78_24_2014}, which could increase ground-state correlations, leading to a rise in excitation energies. However, incorporating this degree of freedom into the PGCM method is complex, and its effect remains to be determined with Gogny energy density functionals.
\section{Summary}
\label{summary}
In this contribution, the calculation of the spectrum of the $N=Z$ nucleus $^{84}$Mo using the PGCM method with the Gogny D1S interaction has been presented. The results indicate that this isotope exhibits multiple shape coexistence, with a triaxially deformed ground state and several low-lying excited $0^{+}$ states with well-distinguished nuclear shapes. Moreover, the experimental data for the ground-state band energies are well described by theory, although part of the quantitative agreement might be incidental due to the limitations of the PGCM implementation used here (interaction adjusted at the mean-field level, and the absence of proton-neutron pairing and/or terms breaking time-reversal symmetry). The inclusion of cranking terms in the calculation and their influence on the spectrum is a work in progress.   
\subsection*{Acknowledgements}
The author acknowledges the Grant PID2021-127890NB-I00 funded by MCIN/AEI/10.13039/501100011033, and the support of the GSI-Darmstadt computer facilities.

\end{document}